\documentclass[11pt,nocontents]{iopart}

\usepackage{caption}
\usepackage[latin1]{inputenc}
\usepackage[T1]{fontenc}
\usepackage[english]{babel}
\usepackage{epsfig}
\usepackage{geometry}
\usepackage{bm}
\usepackage{graphicx}
\usepackage{ulem}
\usepackage{inputenc}
\usepackage{amssymb}
\makeindex 

\begin{document}

\title{Detecting sterile neutrinos with KATRIN like experiments}

\author{Anna Sejersen Riis$^{1,2}$, Steen Hannestad$^1$}
\ead{asr@phys.au.dk, sth@phys.au.dk}
\address{$^1$Department of Physics and Astronomy, University
of Aarhus, Ny Munkegade, DK-8000 Aarhus C, Denmark }
\address{$^2$Institute of Nuclear Physics, Westf{\"a}lische Wilhelms-Universit{\"a}t\\
Wilhelm-Klemm-Str. 9, D-48149 M{\"u}nster}

\begin{abstract}
A sterile neutrino with mass in the eV range, mixing with $\bar\nu_e$, is allowed and possibly even preferred by cosmology and oscillation experiments. If such eV-mass neutrinos exist they provide a much better target for direct detection in beta decay experiments than the active neutrinos which are expected to have sub-eV masses. Their relatively high mass would allow for an easy separation from the primary decay signal in experiments such as KATRIN.
\end{abstract}
\maketitle
\newpage

\section{Introduction}

Currently there is mounting cosmological evidence for an effective number of neutrino species slightly larger than three \cite{Hamann:2010bk,Hamann:2007pi,Hamann:2010pw,GonzalezGarcia:2010un}.
This result will be tested in detail by the Planck mission, and if confirmed will provide solid evidence for physics beyond the standard model.
Additional sterile neutrino states which are partially or fully thermalised via their mixing with the active species is perhaps the most likely candidate for extra relativistic degrees of freedom because thermalisation must occur well below the QCD phase transition temperature. If not, the density of this extra radiation is heavily diluted by the entropy release at the phase transition.

Intriguingly, also in oscillation experiments there is some evidence pointing to the possible existence of sterile neutrinos.
The simplest explanation of the LSND experiment studying $\bar\nu_\mu\to\bar\nu_e$ flavor conversion \cite{Aguilar:2001ty} requires low mass sterile states
\cite{Berezhiani:1995yi,Okada:1996kw,Bilenky:1996rw,Barger:1997yd,Gibbons:1998pg,Barger:1998bn,Bilenky:1998ne,Bilenky:1998dt,Bilenky:1999wz,Peres:2000ic,Giunti:2000ur,GonzalezGarcia:2007ib,Strumia:2002fw,Cirelli:2004cz}
(see also \cite{Goswami:2007su} for a more complete list of
references).

Very recently, the MiniBooNE experiment has released data from a study of the same channel which
seems to confirm the LSND result. However, no compatible excess was found in the $\nu_\mu\to\nu_e$ channel by MiniBooNE. If both results are correct it would require some source of $CP$ -violation. Recent publications suggest two viable options; either there are at least two sterile states \cite{Karagiorgi:2009nb,Sterile2010,Schwetz}, or one sterile state plus non-standard charged current -like neutrino interactions (NSI). These interactions enables a produced (or detected) neutrino to be a linear combination of flavour eigenstates \cite{Schwetzagain}.

In their paper Akhmedov and Schwetz, \cite{Schwetzagain}, find that the 3+1 (active + sterile neutrinos) NSI model gives a somewhat nicer fit than the 3+2 scenario mentioned above. This is especially true when considering global data - which includes the findings of disappearance experiments for $\bar\nu_{e, \mu}$ or $\nu_{e, \mu}$ in addition to the data of the already mentioned appearance experiments with $\bar\nu_\mu\to\bar\nu_e$ \& $\nu_\mu\to\nu_e$. However in the context of beta-decay experiments such as KATRIN \cite{katrin} (which will be the topic of this paper) one measures the number of electrons emitted as a function of their energies. So even though the electron neutrino might become a combination of flavour eigenstates the electron still has a well-defined energy. Thus, as a consequence of simple energy conservation, the electron spectrum is only sensitive to the neutrino mass states and a beta-decay experiment would not be sensitive to the NSI. However, it should be noted that in the presence of NSI the effective mixing angle appearing in the beta spectrum is not the simply $U_{es}$, but rather the contribution of the fourth mass state to the electron neutrino state, as defined in Eq.~2 of Ref.~\cite{Schwetzagain}.

Considering on the other hand the 3+2 model, the two additional low mass sterile states might at first seem incompatible with cosmological
neutrino mass constraints. But as it turns out the mass bound is actually significantly relaxed by the presence of
extra energy density, and it is possible to have sterile states with masses close to 1 eV, compatible with the MiniBooNE result \cite{Melchiorri:2008gq,Acero:2008rh,Dodelson:2005tp}.

Very interestingly in both models the sterile neutrinos would have large mixing with $\bar\nu_e$ such that in nuclear beta decay it might be possible to see direct evidence for the existence of sterile states. Additionally one would not have to worry about $CP$-phases causing cancellations - as would be the case in neutrinoless double beta decay - because the kinematic neutrino mass is an incoherent sum of the mass states.

In this paper we perform a forecast for KATRIN-like experiments in terms of the prospects for observing such sterile states. Even though an active neutrino mass of 1 eV is still allowed by beta decay experiments it is disfavoured by cosmological data (and would require neutrinos to be Dirac particles in order not to have produced observable neutrinoless double beta decays) and we therefore consider a hierarchy in which all three active species are very light, with masses too small to be seen in experiments like KATRIN, while the sterile state(s) have masses in a range accessible by KATRIN (see also e.g.\ \cite{petcov,farzan} for earlier discussions of this possibility). We then investigate how well such an experiment will be able to detect eV mass sterile states, depending on their mixing with $\bar \nu_e$. We wish first and foremost to get an overview of the experimental detection probabilities. Therefore we use a rather large sterile parameter space in combination with simulations that take all the experimental settings of a KATRIN-like experiment into account. The next section contains a description of the analysis framework used, Section 3 is a presentation of our main results - for both a 3+1 and a 3+2 scenario - and finally in Section 4 we make some concluding remarks.

\section{Methodology}

As mentioned we use a realistic simulation of the KATRIN experiment - commissioned to begin data-taking in 2012 - to investigate if a similar experiment will be able to see direct evidence for sterile neutrino states.\\
The main ingredient in the analysis is a toy model Monte Carlo and analysis programme for KATRIN-like experiments written by C.~Weinheimer. The programme contains all the major experimental settings and has previously been used to forecast the experimental sensitivity to the neutrino mass. A much more thourogh simulation procedure is currently being built for the KATRIN collaboration. This will take into account the finer details of the Windowless Gaseous Tritium Source, the differential pumping and cryogenic trapping sections in the transportation of the electrons from the source to the spectrometer, and of course the entire final electrodynamic properties of the spectrometer. Currently however neither the spectrometer or the source have been completed and an ultra-realistic simulation might therefore be subject to final changes anyway. So for the purpose of current forecasts of the detection abilities of a KATRIN-like experiment  this is the most realistic simulation tool one can get. For a thorough description of the workings and parameters of this code see \cite{KAT:04}. We have implemented the $\chi^2$-routine of the programme in a modified version of COSMOMC \cite{COSMOMC} with cosmology turned off and flat priors on all input parameters.

In doing so we have performed a bayesian analysis on the theoretical beta spectra while keeping the error bars of the original Monte Carlo driver \cite{me}. It should be noted that the mixing angles were held fixed in this analysis. While this is of course a perfectly valid way to perform the analysis one may argue that the mixing angles should be free parameters as well as the masses. However the mixing angles  - or more precisely their corresponding entry in the leptonic mixing matrix - determine the amplitude of the electron spectra for each separate neutrino mass state. When looking at the sum of the mass states, that is the total beta-decay spectrum, it is a lot easier to extract knowledge of the neutrino masses (position of the kinks in the spectrum) than the mixing angles (relative height of the kinks). So in order to get well-constrained results we have kept the mixing angles fixed and one can of course perform such an analysis for as fine a grid of mixing angles as desired. Finally, oscillation experiments are much better suited for precise measurements of mixing angles than beta-decay experiments. In a realistic setting a measurement by KATRIN will be combined with short baseline oscillation data to obtain good constraints on both mass and mixing angle.

To the theoretical description of the single neutrino beta spectrum we then add massive sterile neutrinos connected to the active neutrino by mixing angles. We note that KATRIN can in fact not resolve the mass squared differences between the known active states (of $\Delta m_{12}^2=8\times 10^{-5}$ eV$^2$ and $\Delta m_{23}^2=|2.6\times 10^{-3}|$ eV$^2$) so in the following we can safely keep treating the active sector as one neutrino\footnote{Obviously one could also imagine a scenario where both active and sterile species are massive. However we chose to keep the active state(s) massless in this investigation both for simplicity and to satisfy current cosmological bounds as far as possible}.

In complete analogy with Eqs. 16 and 21 of \cite{Masood:07} the total spectrum is a weighted sum of the spectra for the individual mass states. Each mass state is weighted by the relevant mixing angles as explained above. That is the appropriate element of the corresponding leptonic mixing matrix, $U$. In the case of the 3+2 scenario the full 5-neutrino description $U$ would look like this
\begin{equation}
 \centering
U = \left( \begin{array}{ccccc}
U_{e1}& U_{e2} & U_{e3}&U_{e4}&U_{e5} \\
U_{\mu 1}& U_{\mu 2} & U_{\mu 3}&U_{\mu 4}&U_{\mu 5} \\
U_{\tau 1}& U_{\tau2} & U_{\tau 3}&U_{\tau 4}&U_{\tau 5} \\
U_{s_1 1}& U_{s_12} & U_{s_1 3}&U_{s_1 4}&U_{s_1 5} \\
U_{s_2 1}& U_{s_2 2} & U_{s_2 3}&U_{s_2 4}&U_{s_2 5} \end{array} \right),
\end{equation}
and a $4\times 4$ analogy would play the same role in any 3+1 model. For the remainder of this paper the weighting factor is written $|U_{ei}|^2$ ('e' representing the electron neutrino and 'i' representing the contribution of mass state i).

We begin by considering only one sterile neutrino. Under the assumption that the three active neutrinos can be treated as one we investigate firstly $2\times 2$ mixing schemes where one active and one sterile species are considered as two separate blocks connected via a single mixing angle $\theta$. In order to satisfy current mass bounds on the active neutrinos we shall assume the active neutrino to be massless and the sterile neutrino to have a given mass and mixing angle. We do not investigate the reverse mass ordering and we keep the assumption of active massless and sterile massive neutrinos for the investigation of the 3+2 scenario as well. In our $2\times 2$ mixing matrix we must now have $|U_{ee}|^2=1-|U_{es}|^2$, in order to conserve probability.

We investigate the response of a KATRIN-like experiment to the presence of sterile neutrinos for a grid of $11\times 11$ combinations of sterile neutrino masses and mixing weights. As a starting point for our choice of the parameters $m_s$ and $U_{es}$ we look at the analysis of 3+2 scenarios in \cite{Schwetz} and \cite{Goswami:2007su}. They give the following best fit area for 2 additional sterile neutrinos:
\begin{equation}
\Delta m_{s1}^2=6.49 \, {\rm eV}^2 \; \;   |U_{e4}|^2=0.12
\label{eq:best11}
 \end{equation}
 \begin{equation}
\Delta m_{s2}^2=0.89  \, {\rm eV}^2 \; \;   |U_{e5}|^2=0.11,
\label{eq:best12}
\end{equation}
where $|U_{e4}|^2=\sin^2\theta_{(\bar\nu_\mu\to\bar\nu_e),1}$ and $|U_{e5}|^2=\sin^2\theta_{(\bar\nu_\mu\to\bar\nu_e),2}$.
Here we want to investigate the sensitivity of a KATRIN like experiment to a relatively wide range of mass of mixing parameters, including the values above.
We take the sterile mass to be larger than the projected KATRIN sensitivity of 0.2 eV, and to be conservative we investigate masses up to 6.4 eV. We note that such large masses are strongly disfavoured by cosmology if the sterile neutrinos are equilibrated in the early universe. However, even for large mixing angles, equilibration can be blocked by the presence of a e.g.\ a small but non-zero lepton asymmetry and therefore the cosmological mass bound does not necessarily constrain the upper mass range.
Meanwhile $U_{es}$ is varied from $5.5\times 10^{-4}$ up to 0.18, i.e.\ a very wide range of mixing angles. Large values of $U_{es}$ are disfavoured by disappearance data (the most recent analysis of which can be found in \cite{Schwetzagain}), but are included here for completeness.
The exact values of $m_s$ and $U_{es}$ in the $11 \times 11$ matrix of parameters studied are listed in Table \ref{tab:allset1}.

\begin{table}[htb!]
\centering
  \begin{tabular}[htb!]{ | l || c | c | c | c | c | c | c | c | c | c |r | }
    \hline
    $m_s$ [eV] & $0.2$ & $0.28$ & $0.4$ & $0.56$ & $0.8$ & $1.14$ & $1.6$ & $2.28$ & $3.2$ & $4.36$ & $6.4$ \\ \hline
    $|U_{es}|^2$ & $0.00055$ & $0.001$ & $0.0018$ & $0.0033$ & $0.0055$ & $0.01$ & $0.018$ & $0.033$ & $0.055$ & $0.1$ & $0.18$ \\ \hline
    \end{tabular}
    \caption{Sterile neutrino parameters
  \label{tab:allset1}}
  \end{table}

We will be calculating the statistical uncertainty on the neutrino masses. KATRIN's systematical error is currently estimated to be around 0.017 eV, and we shall keep that as our systematic error throughout this investigation. \\
In the standard one-active-neutrino case three factors affect the statistical uncertainty of the neutrino mass. Firstly the signal count rate at the beta-spectrum endpoint. This can be calculated as a combination of the column density of the source, the magnetic fields of the spectrometer and source (these determine the allowed opening angle), and the cross section of the spectrometer. In our case the count rate near the endpoint is controlled via an amplitude factor that is multiplied onto the final spectrum (meaning the spectrum after we take final states into account and convolute it with the experimental response function). Such a factor - which we from now on will call simply the amplitude - depends of course on the experimental settings. In the case of KATRIN those settings would be mainly the source column density, $\rho_d=5\cdot10^{17}$ molecules/cm$^2$, the allowed opening angle, 50.77$^{\circ}$, and the diameter of the effective analysis plane of the spectrometer, 9 m. Given these settings the amplitude has the value 477.5 Hz. For reference we include the full calculation of the amplitude parameter in Appendix A.

The second important issue is the background count rate which can severely obscure potentially interesting features near the endpoint of the spectrum. And thirdly the energy resolution which determines how well we can see the shape of the beta spectrum.

To sum up: The KATRIN experiment has an amplitude of 477.5 Hz, a background rate of of 0.01 Hz and an energy resolution of 0.93 eV. These are the parameters we can later tune to simulate an improved Tritium beta decay experiment. A full list of the parameter values we have investigated can be found in Table \ref{tab:allset2}.

For any experiment we want to calculate the detection potential for a second neutrino, and the statistical uncertainty on the massless component.\\
\begin{center}
\begin{table}[htb!]
\centering
  \begin{tabular}[htb!]{ |l || c | c | c | c | r | }
    \hline
    Amplitude [Hz] & $477.5$ & $750.0$ & $1000.0$ & $1250.0$ & $1500.0$\\ \hline
    Background [Hz] & $0.01$ & $0.005$ & $0.001$ & $0.0005$ & $0.0001$\\ \hline
    Energy resolution [eV] & $0.93$ & $0.5$ & $0.1$ & $0.05$ & $0.01$\\ \hline

    \end{tabular}
    \caption{External parameter settings
  \label{tab:allset2}}
  \end{table}
  \end{center}
\section{Results}

\begin{figure*}[htb!]
\begin{center}
\includegraphics[scale=0.8]{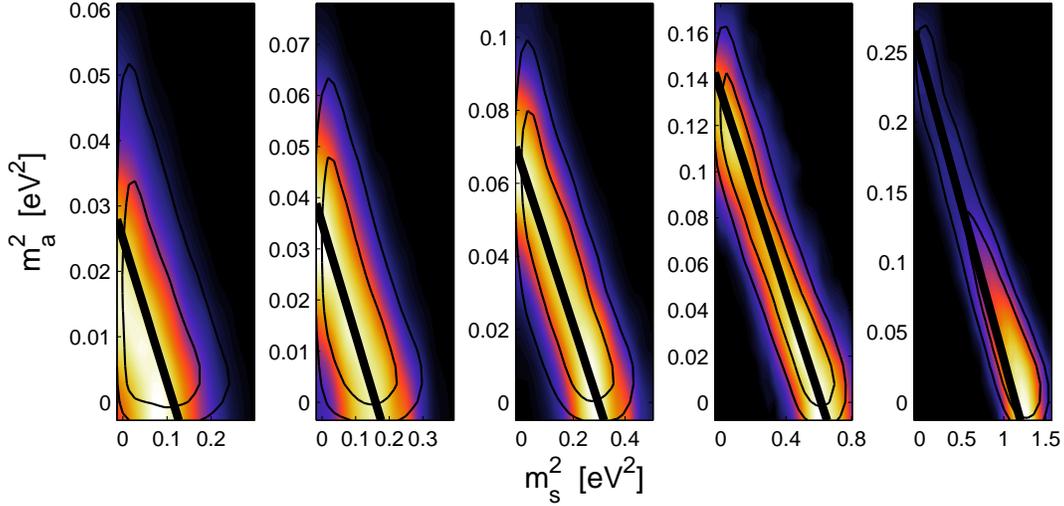}
\end{center}
\caption{The 2D likelihood distributions for the active (y-axis) and sterile (x-axis) neutrino mass-squared for $m_s^2=\{0.0784,\; 0.16,\; 0.3136, \; 0.64,\; 1.2996\}$eV$^2$. In all cases $|U_{es}|^2$=0.18. The color of the figures show how many times the COSMOMC Markov Chain Monte Carlo (MCMC) has probed a specific area of the parameter space and the lines indicate the 1$\sigma$ and 2$\sigma$ likelihood contours. The mass states are clearly anti-correlated for the lowest $m_s^2$ -values. Also shown is the line $m_a^2=m_a^2(m_s^2=0.0)-(0.18/0.82)\cdot m_s^2$. Clearly the two states become interchangeable and dependent on the mass of the other state. This line describes the functional dependence of the contours in the 2D plots fairly well until the mass of the sterile neutrino is so large that the 2 mass states can be separated.
\label{fig:figcont}}
\end{figure*}

Let us start by examining the sigma detection potential for the sterile neutrino component in the standard KATRIN-like setup. Figure~\ref{fig:figa}  in Appendix B shows both the detection potential for the sterile neutrino and the standard deviation on the mass of the active neutrino. As one would expect, effects are clearest in the sterile sector (for which we vary the parameters), but both graphs have a curious feature around $m_s$ = 1.0 eV. At this point in of parameter space the detection potential drops and the standard deviation grows. Inspecting the COSMOMC generated 2D likelihood distributions one can see easily that the two mass states are in fact anti-correlated up to around $\simeq$ 1 eV. An illustration for a mixing weight of 0.18 is shown in Figure~\ref{fig:figcont}. From a simple Taylor expansion of the beta spectrum:
\begin{eqnarray*}
\centering
\frac{dN_{\beta}}{dE_e} & \propto & \sqrt{(E_0-E_e)^2 - m_{\nu}^2}  \\
                        & \propto & (E_0-E_e) - \frac1{2(E_0-E_e)} \cdot (1-|U_{es}|^2) \cdot \left[m_a^2 +\frac{|U_{es}|^2}{1-|U_{es}|^2}\cdot m_s^2\right],
\end{eqnarray*}
we can get the functional shape of the contours in the 2D-plot of Figure~\ref{fig:figcont}. Here $E_e$ is the electron energy and $E_0$ is the theoretical Q-value (with neutrinos assumed to be massless). With the $\left[m_a^2 +\frac{|U_{es}|^2}{1-|U_{es}|^2}\cdot m_s^2\right]$-term representing the total neutrino mass it is clear that one can write the following expression for the active mass:

\begin{equation}
\centering
m_a^2=m_{tot}^2-(0.18/0.82)\cdot m_s^2=m_a^2(m_s^2=0.0)-(0.18/0.82)\cdot m_s^2
\end{equation}
When both masses are very low the mass states act in exactly the same manner in the spectrum and the $\chi^2$-function cannot tell them apart. However for high enough masses, when the Taylor expansion is no longer valid, the situation changes. Then the mass states are not as entangled and the well known correlation between the beta spectrum endpoint (which is a free parameter in the analysis - see \cite{KAT:04} for further explanations) and the neutrino mass comes into play. That is, higher values of $E_0$ means higher values of $m_a^2$. Because of the mixing with the sterile mass state, higher values of $E_0$ also mean higher values of $m_s^2$ and we get $m_a^2 \propto|U_{es}|^2 \cdot m_s^2$. This rather weak correlation is the cause of the increase in $\sigma_{stat}(m_a^2)$ for high $m_s^2$ and high $|U_{es}|^2$ that can be seen in Figure~\ref{fig:figa}.

But of course the $\chi^2$-function also depends on amplitude and background. And so it turns out that the first point of low sensitivity moves towards lower values of $m_s$ when we start tuning the experimental configuration. In other words, the signal to noise ratio is a decisive factor here.

While trying to understand what else might affect this situation one should bear in mind that the theoretical $\beta$-decay spectrum is folded with the electronic spectrum of molecular Tritium and with KATRINs response function (describing the energy loss of electrons in the source and the transmission function of the spectrometer). Those effects are already included in the code so that we may take them into account. Thus the problem should not be the many convoluted functions. However, another potentially problematic point of the analysis is the fact that the spectrum is a continuous function but the measurements are performed at discrete energy intervals around the theoretical Q-value, $E_0$. Actually the measurement time distribution has been optimized for the standard one-neutrino case in which the best sensitivity to the neutrino mass is reached when the signal strength equals twice the background count rate - see \cite{Kraus:05}. We have omitted this rather complex optimization for our sterile neutrino scenario and that could be a part of the problem. There are certainly hints that some of the discontinuities in our results may be caused by numerical bad points stemming from the measurement time distribution.

As it is, our results show that for the current settings the sensitivity of a KATRIN-like experiment to both neutrino states is lowered unless the mass difference is sufficiently large.

Beyond the signature at 1 eV the graphs shows the basic and expected result that a 3$\sigma$ detection (above the red mesh) of the relic neutrino requires rather high values of mixing weight and (or) mass. It also shows that for most of the parameter range having a second neutrino in the mix actually improves the statistical uncertainty on the mass-squared of the active neutrino. This happens mostly when the extra dimension of parameter space due to the small admixture of sterile neutrino absorbs a lot of the uncertainty, resulting in a low $m_s$ detection probability.

\begin{figure*}[htb!]
\centering
\includegraphics[scale=0.55]{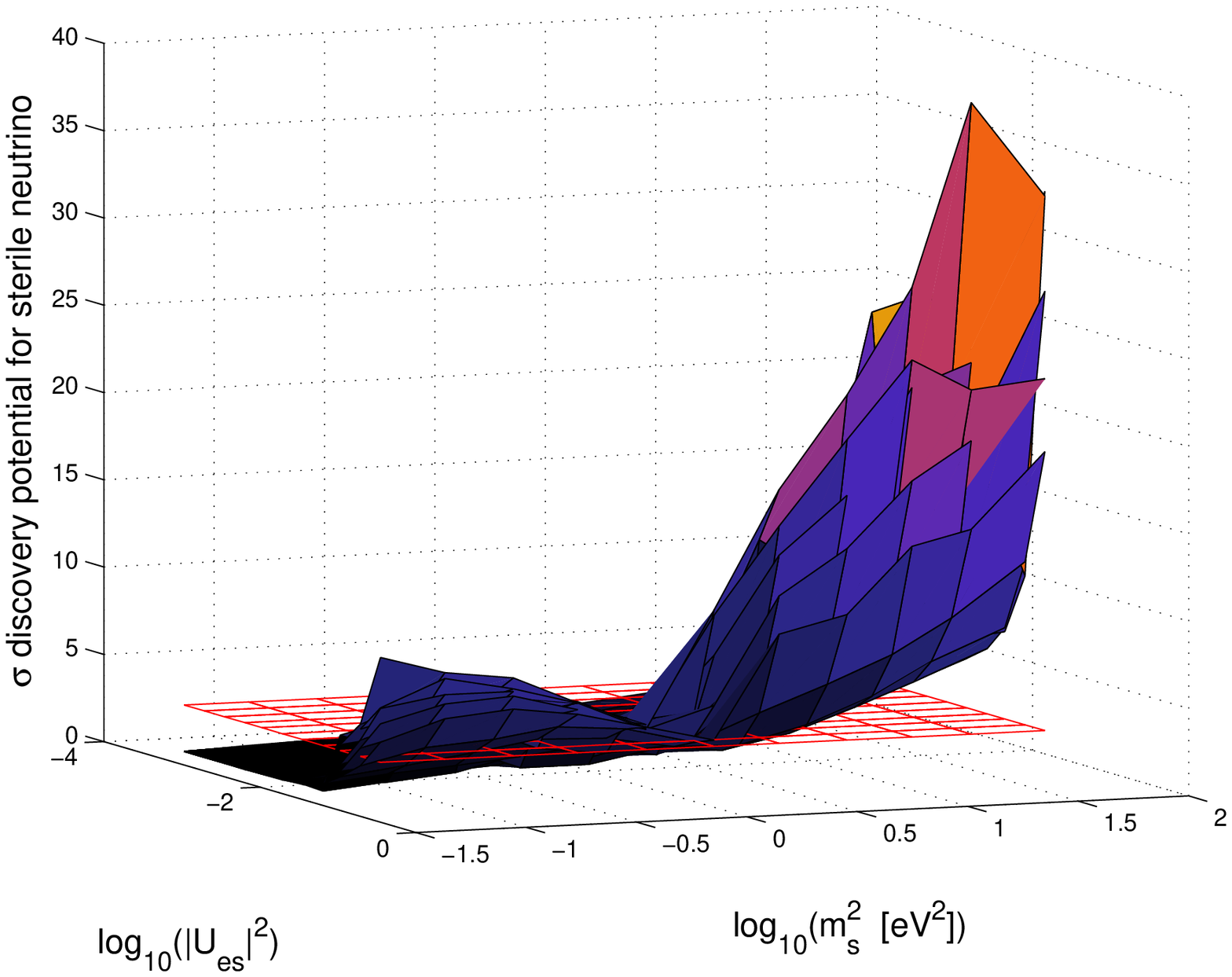}
\vspace{0.01cm}
\includegraphics[scale=0.55]{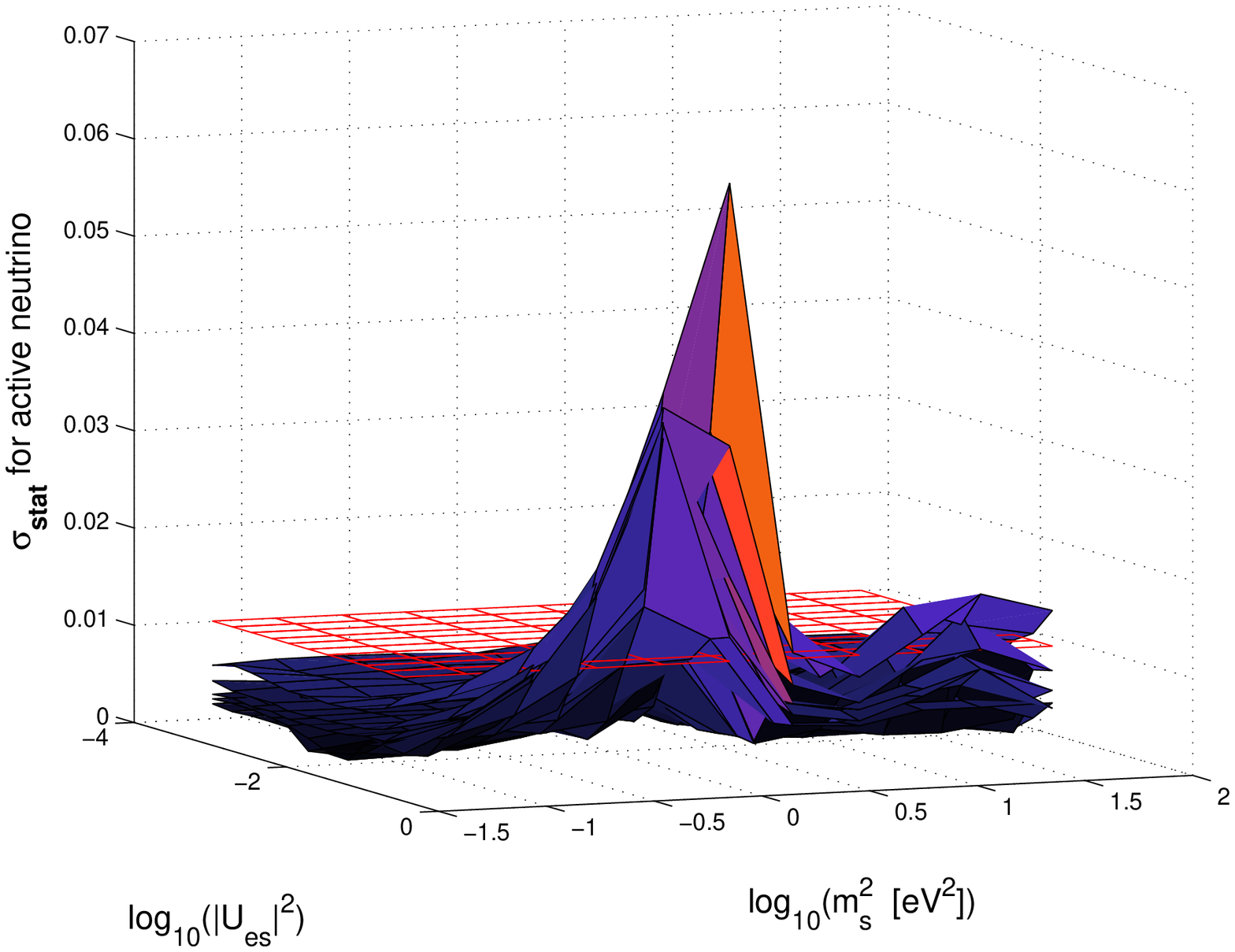}
\caption{The upper figure depicts a comparison of all the amplitude plots for the sterile neutrino while the lower figure shows the corresponding comparison for the active neutrino component. Included in these graphs is the standard case (being respectively the lowest sheet in the upper figure and the highest sheet in the lower figure). In both figures one can see the low sensitivity point move to slightly lower mass values as the amplitude rises. The sensitivity to active mass component gets steadily better as one can see from the nice regular sheets in the low $m_s^2$, $|U_{es}|^2$ area of the lower plot. The improvement on the sensitivity to the sterile neutrino component is not a particularly smooth function of the amplitude suggesting (again) numerical bad points in the grid.
\label{fig:figampste2}}
\end{figure*}

If we enhance the amplitude of the experiment one can clearly see the effect (Figure~\ref{fig:figampste2}). However almost nothing happens for simultaneously low values of $|U_{es}|^2$ and $m_s$ as compared to the standard case. Given high values of the sterile parameters there is quite some development. But it is certainly not a very smooth development as a function of the rising amplitude. This could again hint at numerical discrepancy between the 'real' and the measured value of certain parameters as a consequence of the measurement time series. In Figure~\ref{fig:figampste2} one can also see how the statistical uncertainty on the active neutrino gets better for each step - barring the point of low sensitivity. This effect definitely corresponds with predictions. By looking closely at the figures it can also be made out that the point of low sensitivity moves slightly toward lower mass values as the amplitude grows.

The second important factor is the background count rate. The effect of a lower background is shown in Figure~\ref{fig:figbagste2} of Appendix B and to some extent resembles the case for larger amplitude. The development is somewhat smoother and the effect not so large, but again there is an enhancement of the detection potential for high values of the sterile neutrino components. And as before the lower background has an improving effect also on the statistical uncertainty of the active neutrino mass.

\begin{figure*}[htb!]
\centering
\includegraphics[scale=0.55]{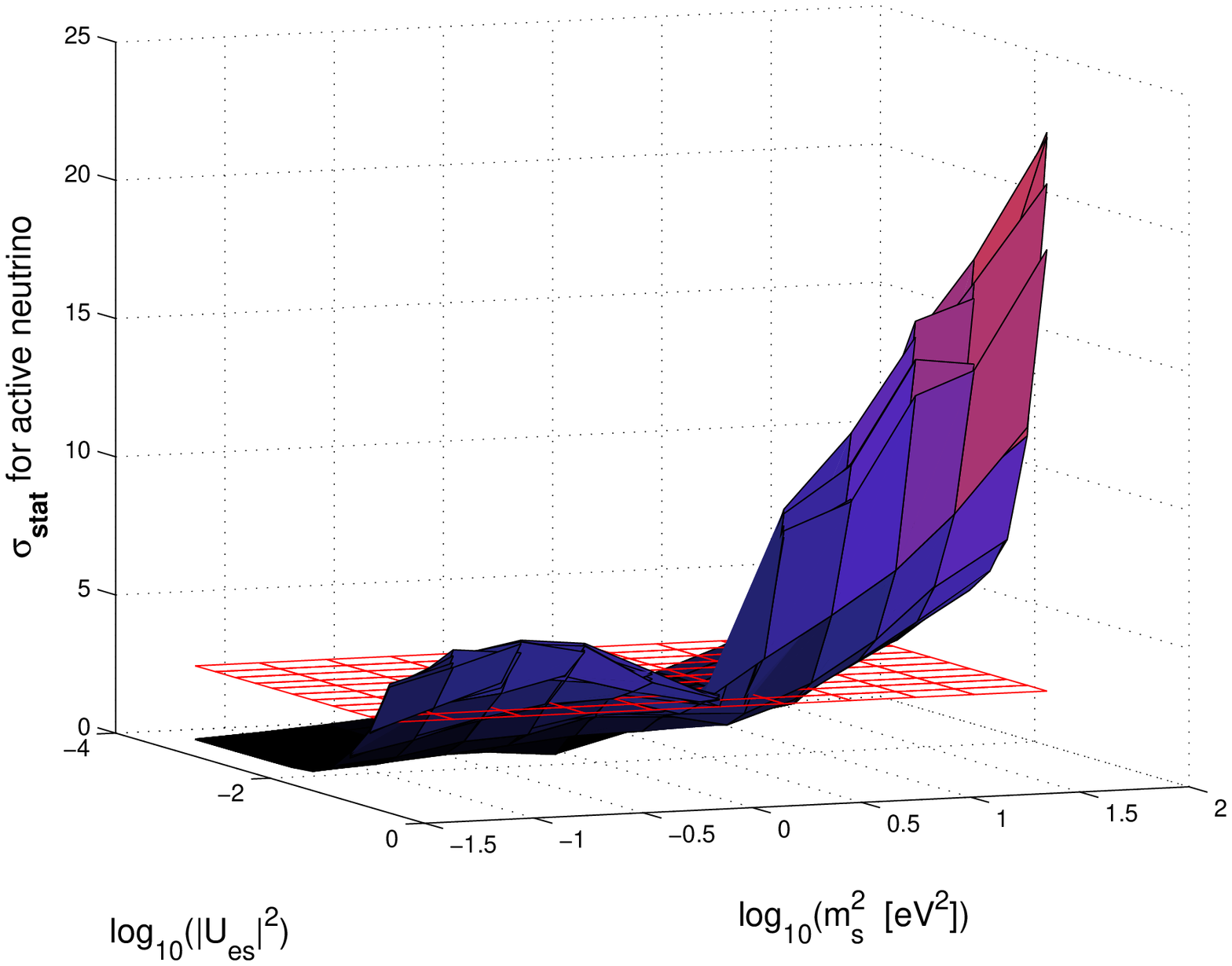}
\vspace{0.01cm}
\includegraphics[scale=0.55]{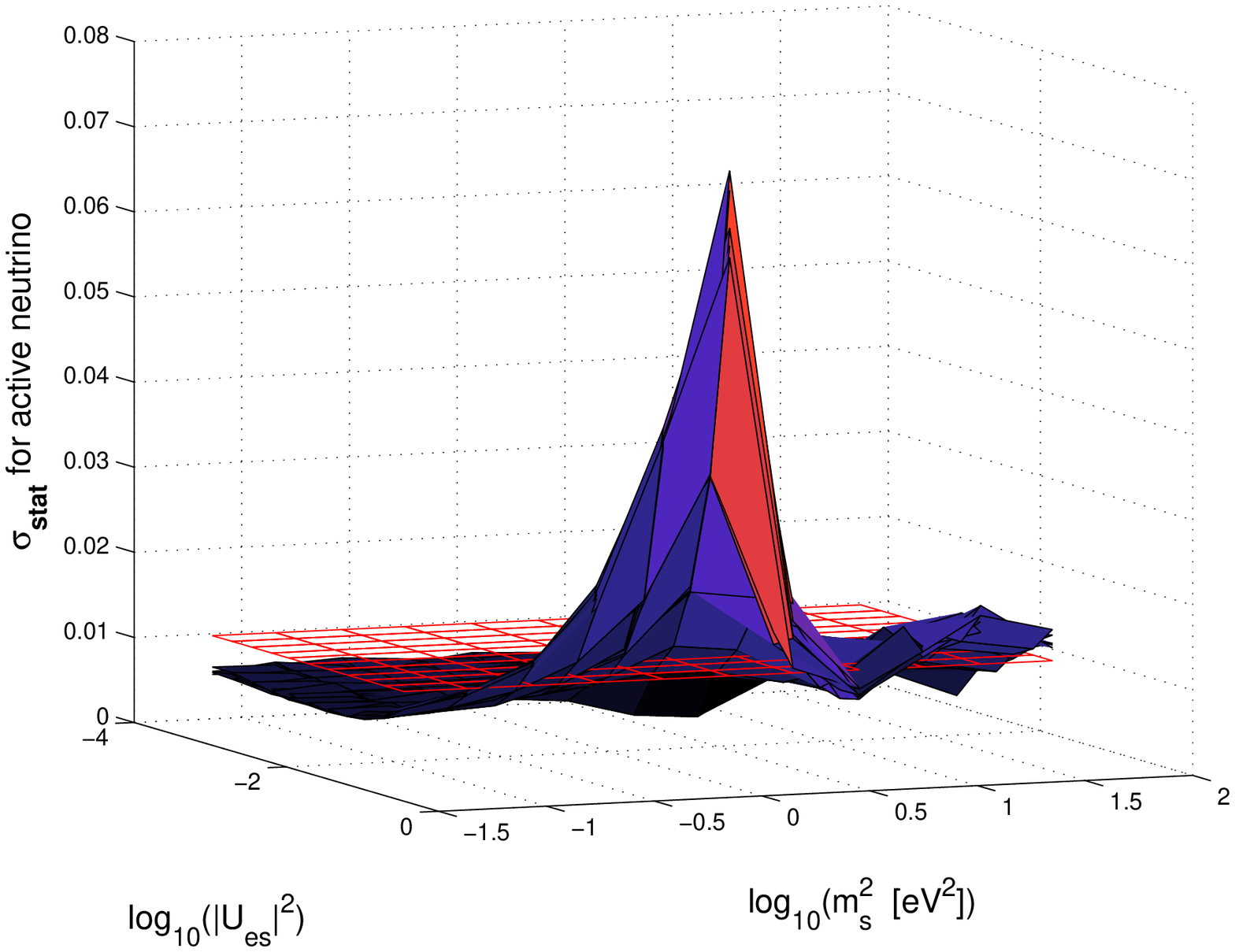}
\caption{The upper figure depicts a comparison of all energy resolution plots for the sterile neutrino mass component. In the sterile component there are some slight changes where $|U_{es}|^2$ is high. In the lower plots one see that the statistcal uncertainty on the active neutrino component hardly changes at all. As before the standard case is included.
\label{fig:figres}}
\end{figure*}

Finally if we try to enhance the energy resolution of the experiment, almost nothing happens. Looking at Figure~\ref{fig:figres} there are only tiny differences in the detection potential for the sterile neutrino as compared to the standard case (again only for high $m_s^2$ and $|U_{es}|^2$), and apart from that not even the statistical uncertainty on the active neutrino mass responds very much. The reason that we get so little effect from an improved energy resolution is the molecular Tritium final states. The width of the final state distribution is currently comparable to the energy resolution (FWHM $\simeq$ 0.7eV) and so one would have to first avoid that smear of the spectrum (i.e. use an atomic source) in order to really gain anything from a better resolution.

\begin{figure*}[htb!]
\centering
\includegraphics[scale=0.7]{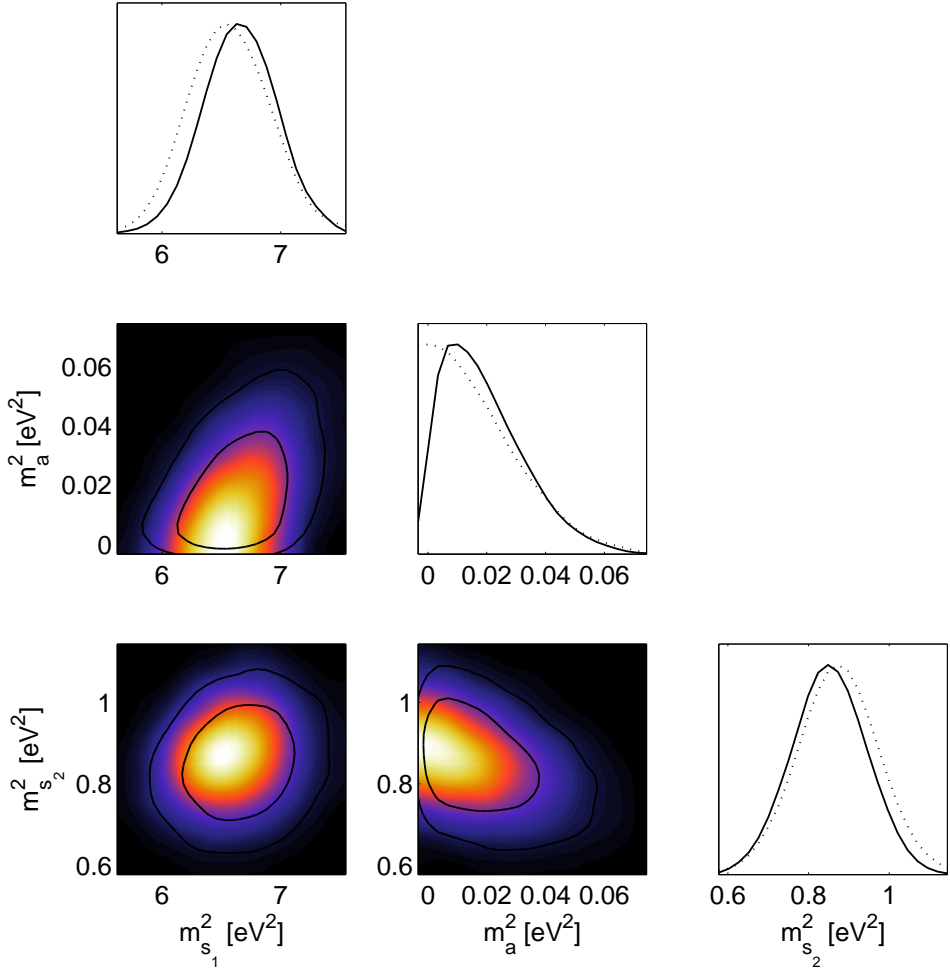}
\caption{The analysis of a 3-neutrino case using the parameter values of Eq.'s ~\ref{eq:best11} and ~\ref{eq:best12}. The contours (dotted lines) mark the likelihood function in the 2D (1D) -distributions, while color (full lines) indicate the data point distribution. The results have converged very well on the input values.
\label{fig:trip}}
\end{figure*}

Let us now take a look at an expanded toy model with two massive sterile neutrinos to see if another added mass state has any damaging effect on our conclusions for the detection abilities of KATRIN-like experiments. Again assuming the active states can be described with one single neutrino the 3+2 scenario gives us four possible mass orderings (the sterile neutrinos being heavier or lighter than the active state, or one light sterile neutrino one massive active neutrino and one more massive sterile neutrino). As before the case of two heavy sterile neutrinos and a massless active state produces the lowest sum of masses and would therefore be most compatible with cosmological bounds. We did a COSMOMC analysis for this model using the best fit point of Eq.'s ~\ref{eq:best11} and ~\ref{eq:best12}. The results are shown in Figure~\ref{fig:trip}.\\
In the analysis we fixed the values of the mixing angles and again let COSMOMC analyse a theoretical KATRIN-like spectrum albeit with Monte Carlo generated errorbars.

The output values of our analysis are\footnote{The means are taken from the best fit sample model and the standard deviations from the marginalization procedure}.
\begin{eqnarray*}
  m_{s1}^2  & = & 6.47 \pm 0.30 \, {\rm eV} ^2 \\
  m_{s1}^2  & = & 0.88 \pm 0.090  \, {\rm eV} ^2 \\
  m_{a}^2  & = & 0.005 \pm 0.014  \, {\rm eV} ^2, \\
 \end{eqnarray*}
in agreement with the input values. We observe that our KATRIN toy model experiment are certainly able to see also two sterile mass states if such a scenario is in fact realized in nature.

\section{Conclusions}

We can see from these investigations that for the highest coupling strengths and masses our KATRIN model will definitely be able to separate one or more sterile neutrino components from the active neutrino component, if they do in fact have a mass and mixing angle in the range proposed by \cite{Sterile2010}, \cite{Schwetz}, \cite{Schwetzagain}, and \cite{Goswami:2007su}. However even if just one of the parameters are large there is a very good chance of detection.

\begin{figure*}[htb!]
\centering
\includegraphics[scale=0.3]{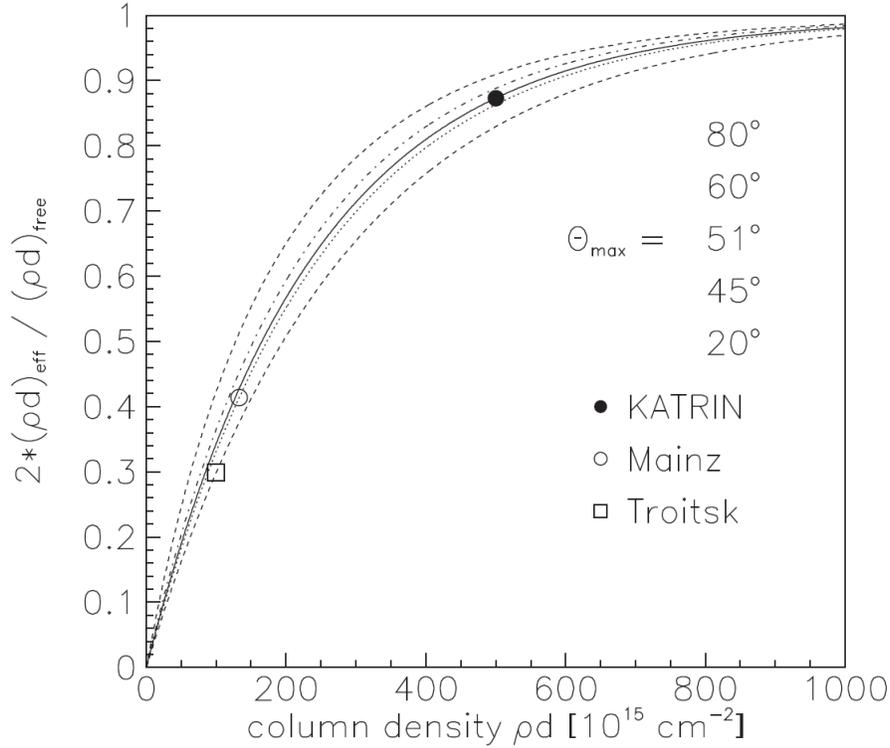}
\caption{The ratio of effetive to free column density as a function of the source column density in the currently proposed source. Clearly the ratio starts saturating at around $\rho_d\simeq $5$\cdot 10^{17}$ molecules pr cm$^2$. The symbols depict the settings for the Troitsk, Mainz and KATRIN experiments, and the lines indicate different maximum allowed starting angles for the spectrometer. KATRIN has $\theta_{max}$=51$\textdegree$. This figure is taken from the KATRIN Design Report \cite{KAT:04}.
\label{fig:sat}}
\end{figure*}

If one wishes to improve the current detection probabilities, the safest way to go is by enhancing the amplitude. However it is a rather complex operation to do so. If one for instance tries to simply increase the amount of source material in KATRIN there will actually not be much to gain. This is because the amount of Tritium has been optimized for the size of the source which in turn has been optimized for the largest realistic spectrometer-size at the time of construction. Put differently; adding more Tritium to the source would not increase the effective column density linearly - one would only see a further saturation as illustrated in Figure~\ref{fig:sat}. If one then constructed a source with a larger cross section one would also need a larger spectrometer. Consequentially - if one used the same strategy as with the current 10 by 25 m spectrometer - certain smaller german villages would have to be rearranged in order to transport such an apparatus to Karlsruhe.

Another promising option is a lowering of the background count rate. Given the numerous background sources in the KATRIN experiment and the continued occurence of new ones this may not be feasible either. Obviously the KATRIN collaboration is making an enormous effort to deal with this problem and maybe when the time comes for data-taking it will be easier to assess the opportunities for a lowering of the background.

Finally improved energy resolution will do very little to help these particular investigations. However the chances of improving the experimental parameters may be better in other future beta decay experiments e.g. MARE or the Project 8 proposal \cite{Monreal:09, Monfardini:06}.

For low values of both sterile parameters it looks like not much can be done to enable a detection of sterile neutrinos with such properties. At least not within a few orders of magnitude of the tunable experimental parameters.

So in conclusion our results show that given the existence of massive sterile neutrinos coupled to the electron anti-neutrino, with the parameters as in Eq.'s ~\ref{eq:best11} and ~\ref{eq:best12}, KATRIN should definitely be able to see them. In fact for a large enough coupling, KATRIN should be able to see a sterile neutrino of any of the mass states we have used. Likewise, for a state massive enough, KATRIN would be able to see the sterile neutrino for any of the mixing angles we have used. Specifically, in our $2\times 2$ mixing scheme the experiment could perform a 3$\sigma$ detection of any of the mass states for $|U_{es}|^2 \gtrsim$ 0.055. Likewise, for $m_s \gtrsim$ 3.2 eV a 3$\sigma$ detection could be made for any of the mixing angles. In case of 1$\sigma$ detection $|U_{es}|^2$ must be $\gtrsim$ 0.018 and $m_s \gtrsim$ 0.8 eV

Furthermore it is worth repeating that because beta-decay experiments measure the incoherent sum: $m^2(\nu_e)=\displaystyle\sum\limits_{i=0}^n |U_{ei}^2| m_i^2$, any $CP$ phases causing the anti-neutrino/neutrino asymmetry in MiniBooNE results will not be able to cancel the potential signal in beta-decay experiments. Therefore KATRIN is in fact an ideal tool for making a direct observation of sterile neutrinos with the properties described in this paper.

\section{Acknowledgements}

The authors would like to thank Christian Weinheimer for enlightening discussions on the subject. We further acknowledge use of computing resources from the Danish Center for Scientific Computing (DCSC). Anna Sejersen Riis was supported by BMBF under contract number 05A08PM1 whilst working on the majority of the research in this paper

 \newpage

\section*{Appendix A: Calculating the 'amplitude' for a KATRIN-like experiment}
The count rate near the endpoint of the Tritium beta spectrum can be written:
\begin{eqnarray*}
\frac{dN_{\beta}}{dt}(E_e) & = & N_{tri}\cdot \varepsilon_{tot} \cdot K \cdot F(E_0,Z) \cdot p_e \cdot (E_e+m_e) \\
&\times &\displaystyle\sum\limits_{i=0}^n W_i (E_0 - V_i - E_e) \sqrt{(E_0 - V_i - E_e)^2 - m_{\nu}^2}.
\end{eqnarray*}
Here $p_e$, $E_e$, and $m_e$ are the momentum, kinetic energy and mass of the electron and $m_{\nu}$ is the mass of the neutrino. $W_i$ and $V_i$ are the probability and excitation energy of the electronic final states. We will now go through the remaining factors of the expression one by one: $N_{tri}$ is the number of Tritium molecules that can be seen by the spectrometer:
\begin{equation*}
\centering
N_{tri}= \rho_d \cdot A_S \cdot 2 \cdot \varepsilon_{tri}.
\end{equation*}
In this equation $\rho_d$ is the source column density, $\varepsilon_{tri}$ is the Tritium fraction in the source (that is the source efficiency) and $A_S$ is the cross section of the source. The relation between $A_S$ and the spectrometer cross section is:
\begin{eqnarray*}
\centering
A_{S} & = & A_{pinch} \cdot B_{max}/B_S  \\
      & = & A_A^{eff} \cdot B_A/B_{max} \cdot B_{max}/B_S \\
      & = & A_A^{eff} \cdot B_A/B_{max} \cdot 1/ \sin^2 (\theta_{max}),
\end{eqnarray*}
where $A_A^{eff}$ is the effective cross section of the spectrometer at the position of the analysis plane. $B_S$, $B_{max}$ and $B_A$ are the magnetic fields of the source, pinch and analysis plane respectively and $\theta_{max}$ is the maximally allowed opening angle. Keeping the dependence on magnetic fields and $\theta_{max}$ in our expressions we now have:
\begin{equation*}
\centering
N_{tri}= \rho_d \cdot A_A^{eff} \cdot B_A/B_{max} \cdot 1/ \sin^2 (\theta_{max}) \cdot 2 \cdot  \varepsilon_{tri}.
\end{equation*}
Next up, the total efficiency or acceptance of the experiment, $\varepsilon_{tot}$, is a combination of the solid opening angle, response function and detector efficiency:
\begin{eqnarray*}
\varepsilon_{tot} & = & \Delta\Omega /4 \pi \cdot f_{res}(E_e,E_0)\cdot \varepsilon_{det}  \\
                  & = & 0.5 \cdot (1-cos(\theta_{max})) \cdot f_{res}(E_e,E_0) \cdot \varepsilon_{det}.
\end{eqnarray*}
Here $f_{res}$ is the normalized response function and $\varepsilon_{det}$ is the detector efficiency.

Finally the factors $K$ and $F(E_0,Z)$ is a collection of constants and the Fermi Function.
\begin{eqnarray*}
K & = & G_F^2 \cos^2(\theta_C)|M|^2/(2 \pi^3 (h/ 2 \pi)^7) \\
  & = & G_F^2 \cos^2(\theta_C)(5.55 \; h/(h/2\pi)^6)/(2 \pi^3 (h/ 2 \pi)^7),
\end{eqnarray*}
with $G_F$ the Fermi constant and $\theta_C$ the Cabibbo angle. The value of the Tritium matrix element, $|M|^2$, can be found in e.g. \cite{Robertson}.

\begin{table}[h!]
\centering
  \begin{tabular}{ |l | r | }
    \hline
    Parameter & Value \\ \hline
    K & $1.76 \cdot 10^{-17}\; \mathrm{s}^{-1} \; \mathrm{keV}^{-5}$  \\ \hline
    $F(E_0,Z)$ & 1.19  \\ \hline	
    $\rho_d$ & $5\cdot10^{21}$ m$^{-2}$ \\ \hline	
    $A_A^{eff}$ & $(\pi \cdot 4.50^2)=64$ m$^2$\\ \hline	
    $B_S$ & 3.6 T\\ \hline	
    $B_{max}$ & 6.0 T\\ \hline	
    $B_A$ & 0.0003  T\\ \hline	
    $\theta_{max}$ & 50.77 $^\circ$ \\ \hline	
    $\varepsilon_{det}$ & 0.9 \\ \hline
    $\varepsilon_{tri}$ & 0.95  \\ \hline
    \end{tabular}
    \caption{Selected KATRIN parameters. More details can be found in \cite{KAT:04}
  \label{tab:allset3}}
  \end{table}

\noindent{In} total we have
\begin{eqnarray*}
\frac{dN_{\beta}}{dt}(E_0) & = & \rho_d \cdot A_A^{eff} \cdot B_A/B_{max} \cdot 1/ \sin^2 (\theta_{max}) \cdot 2 \cdot \varepsilon_{tri}  \\
                          & \times & 0.5 \cdot (1-\cos(\theta_{max})) \cdot f_{res}(E_e,E_0) \cdot \varepsilon_{det}\cdot K \cdot F(E_0,Z)   \\
                          &\times & p_e \cdot (E_e+m_e)\displaystyle\sum\limits_{i=0}^n W_i (E_0-V_i-E_e) \sqrt{(E_0-V_i-E_e)^2- m_{\nu}^2}
\end{eqnarray*}
and
\begin{eqnarray*}
\frac{dN_{\beta}}{dt}(E_0) & = & \mathrm{Amplitude} \cdot \mathrm{keV}^{-5} (1-\cos(\theta_{max})) \cdot f_{res}(E_e,E_0)  \cdot p_e \cdot (E_e+m_e)\\
                          &\times & \displaystyle\sum\limits_{i=0}^n W_i (E_0-V_i-E_e) \sqrt{(E_0-V_i-E_e)^2- m_{\nu}^2}.
\end{eqnarray*}
\noindent{So}, in conclusion our definition of the amplitude is
\begin{equation*}
\centering
\mathrm{Amplitude}=  \mathrm{keV^5} \cdot K \cdot F(E_0,Z) \cdot \rho_d \cdot A_A^{eff} \cdot B_A/B_{max} \cdot 1/ \sin^2 (\theta_{max}) \cdot \varepsilon_{det} \cdot \varepsilon_{tri}.
\end{equation*}
With the parameter values given in Table~\ref{tab:allset3} we get an amplitude of $\underline{477.5\,\mathrm{Hz}}$

\newpage

\section*{Appendix B: Figures}
\begin{figure*}[h!]
\centering
\includegraphics[scale=0.45]{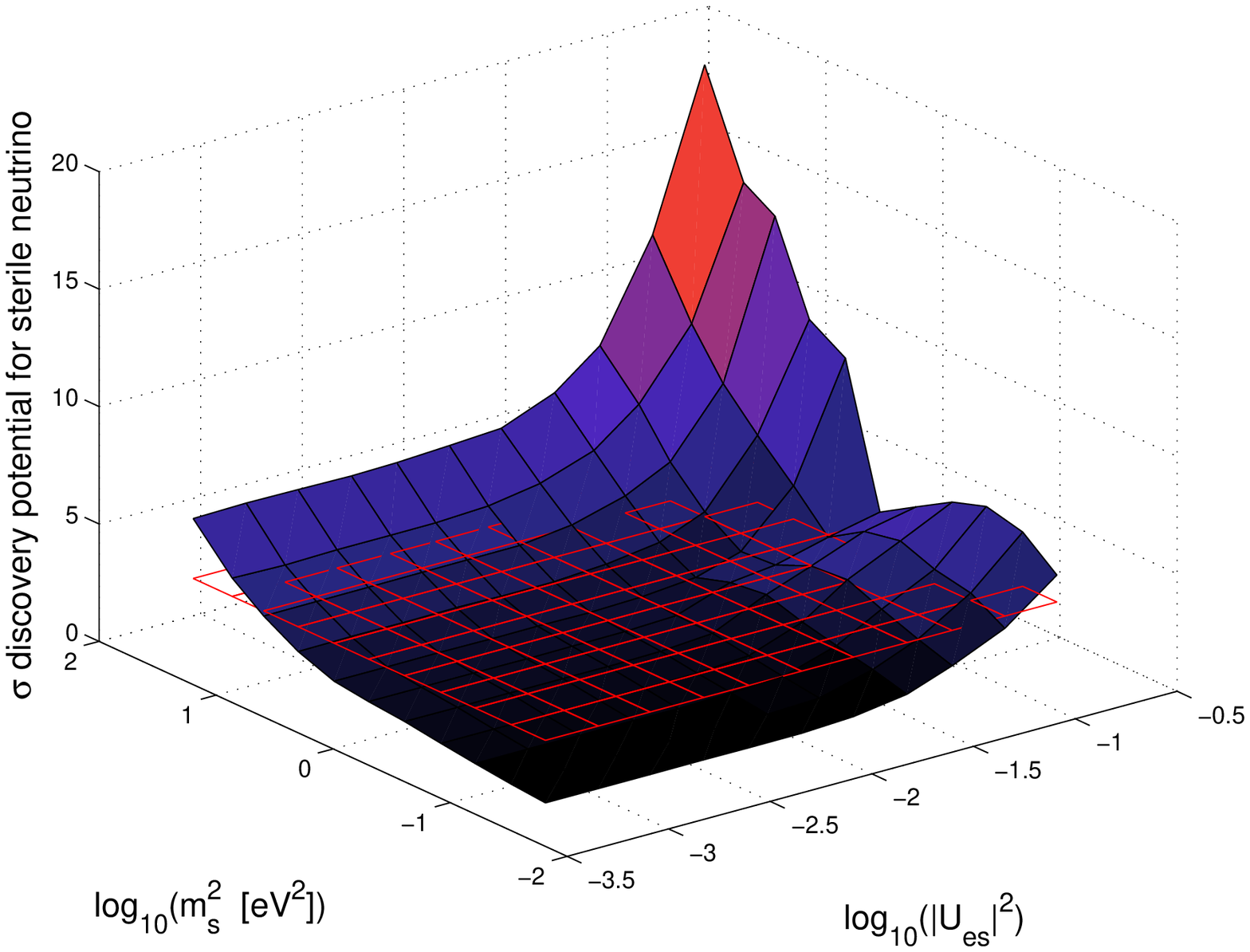}
\hspace{0.01cm}
\includegraphics[scale=0.45]{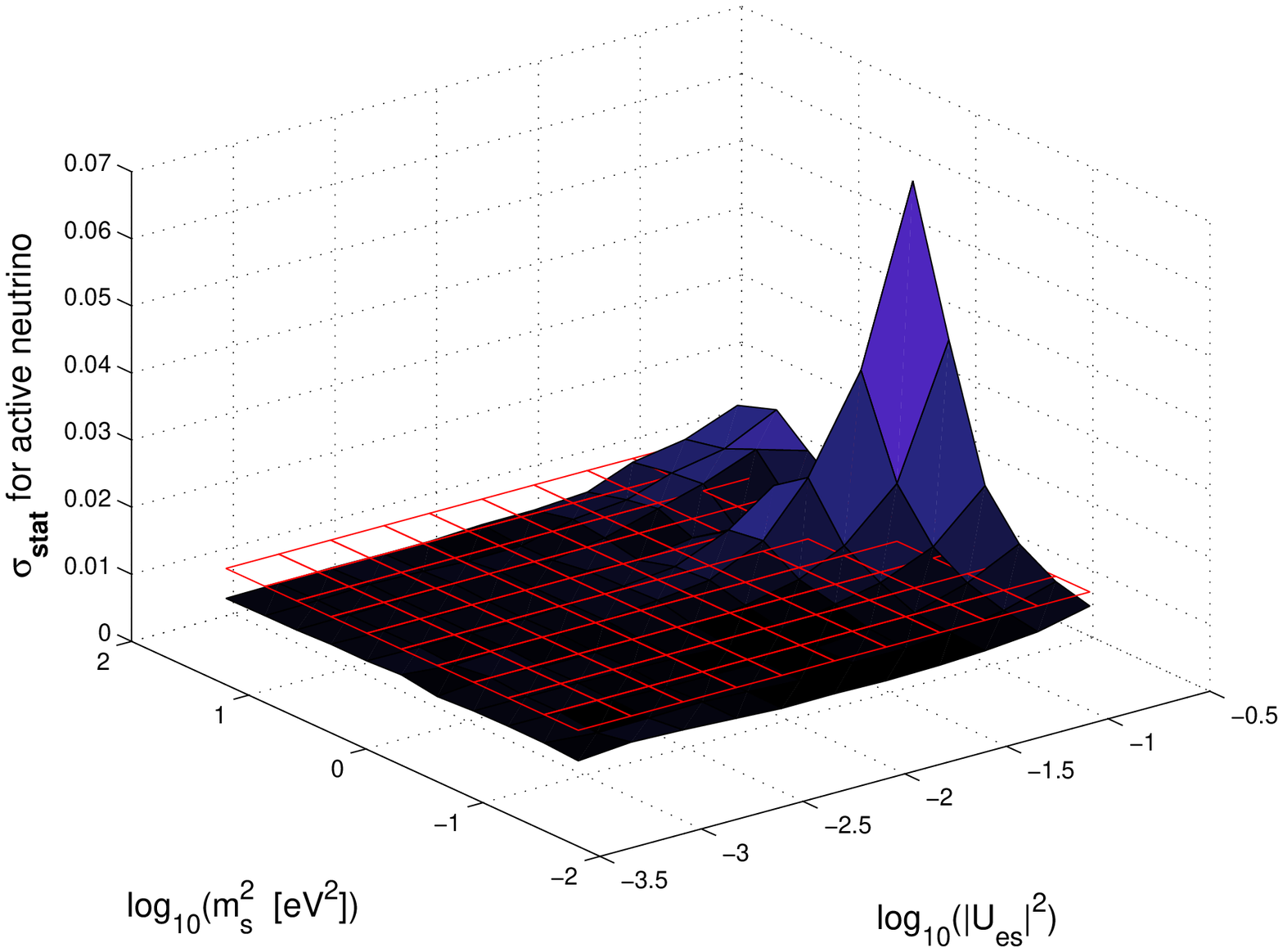}
\caption{The upper figure shows the sigma detection potential of the massive sterile neutrino for the standard KATRIN-like settings, while the lower shows the corresponding statistical deviation (in eV$^2$) on the massless neutrino. The x- and y-axis depicts the logarithm of the sterile mass squared and the mixing weight. The red mesh illustrates the 3$\sigma$ level and the standard one-neutrino statistical uncertainty of around 0.012 eV$^2$ respectively (for this analysis method). As one would expect the mass and the mixing weight must be rather high in order to get a good detection of the sterile component.
\label{fig:figa}}
\end{figure*}

\begin{figure*}[htb!]
\centering
\includegraphics[scale=0.55]{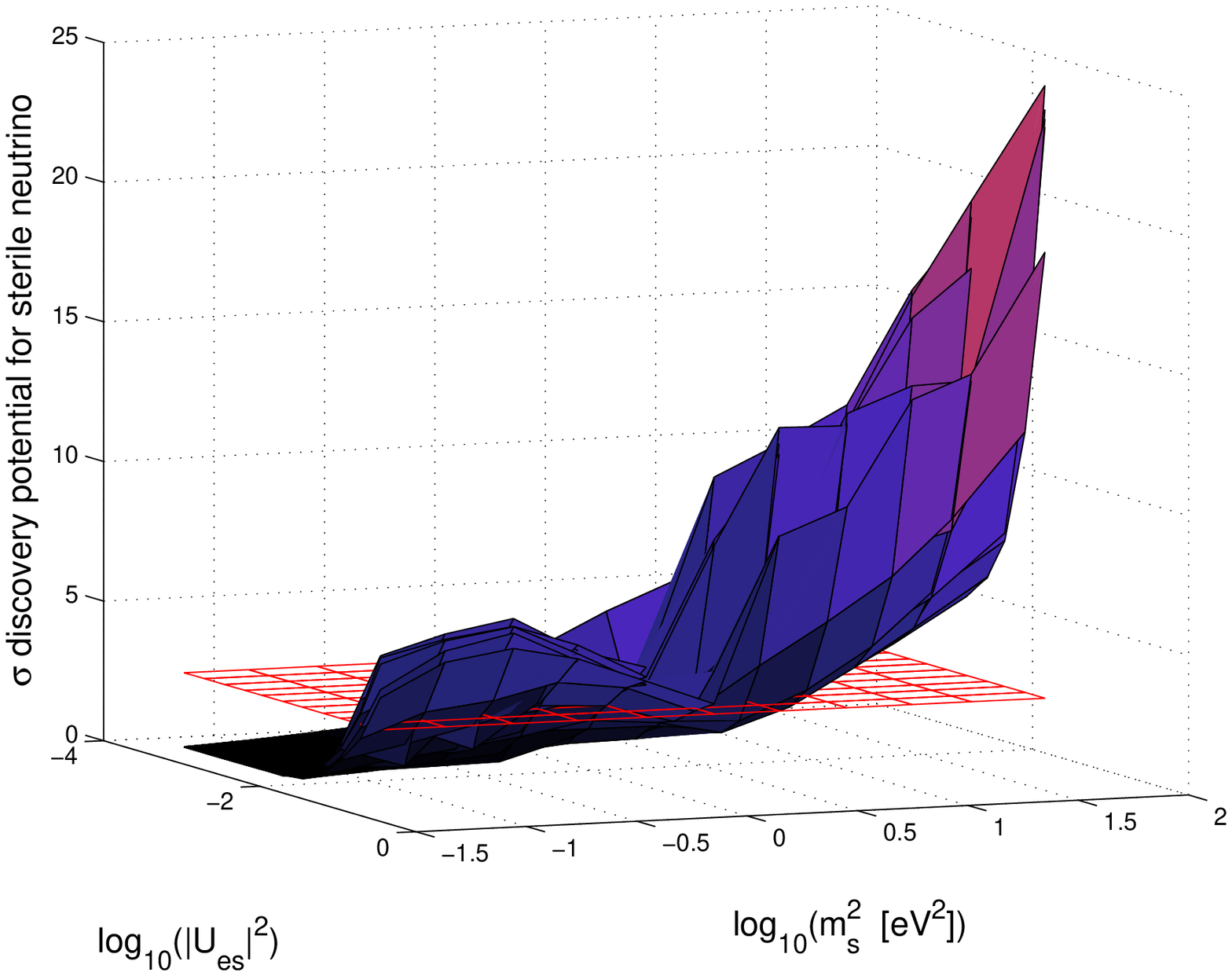}
\vspace{0.01cm}
\includegraphics[scale=0.55]{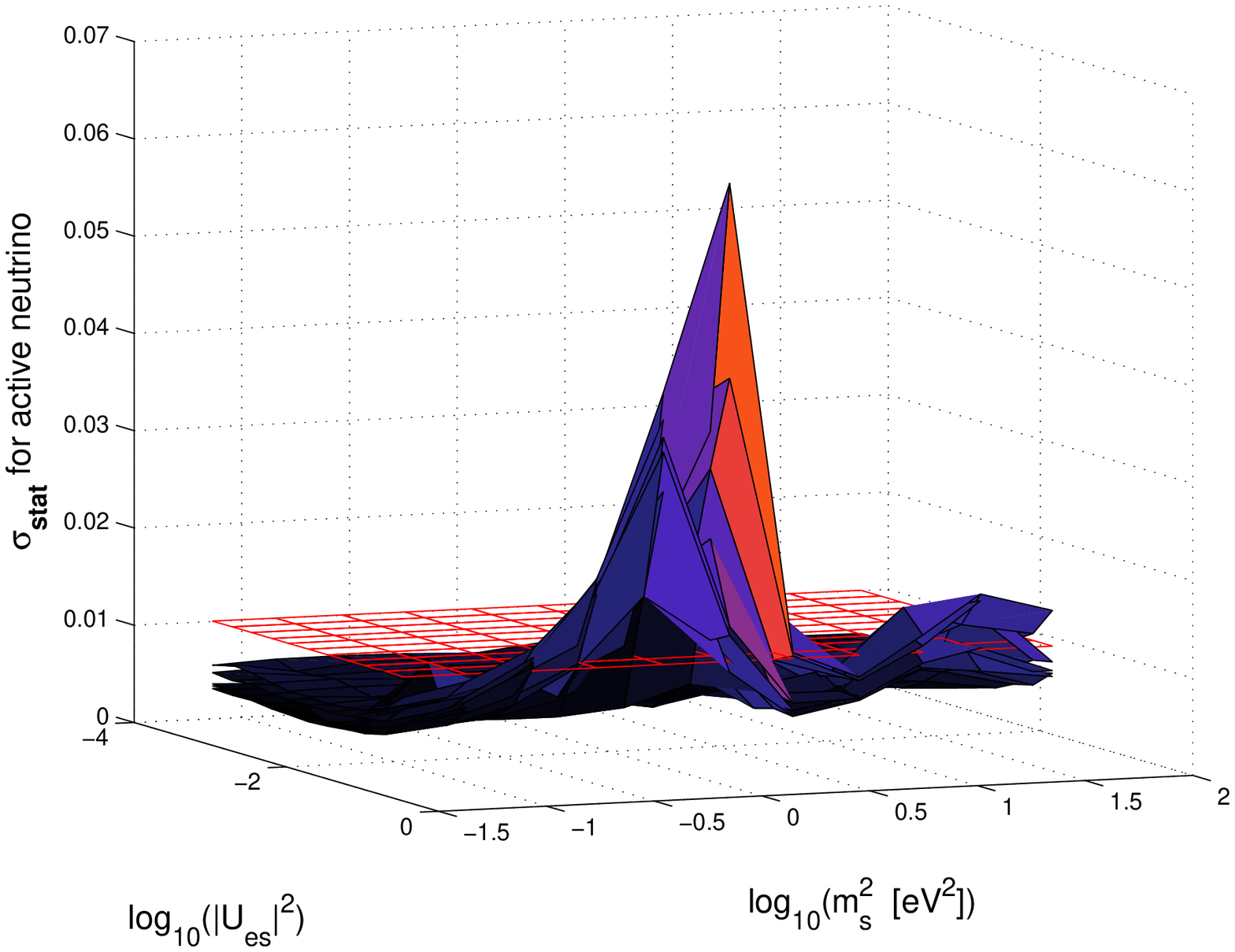}
\caption{The upper figure depicts a comparison of all background plots for the sterile neutrino. One sees slight improvements on the sensitivity to the sterile neutrino component and mostly in the high $|U_{es}|^2$ limit. The lower figure shows the comparison for the active neutrino component. In both cases one can see the low sensitivity point move to slightly lower mass values as the background is lowered. Also we again see the stable lowering of the statistical uncertainty on the active neutrino mass component. Included in these graphs is also the standard case.
\label{fig:figbagste2}}
\end{figure*}

\end{document}